\documentclass[a4paper,11pt]{article}
\pdfoutput=1 

\usepackage{jinstpub} 
\usepackage{caption}
\usepackage{subcaption}
\usepackage{lineno}

\title{Upgrade of the CMS Resistive Plate Chambers for the High Luminosity LHC}
\author[a,1]{A. Samalan,\note{Corresponding author.}}
\author[a]{M. Tytgat,}
\author[b]{G.A. Alves,}
\author[b]{F. Marujo,}
\author[c]{F. Torres Da Silva De Araujo,}
\author[c]{E.M. Da Costa,}
\author[c]{D. De Jesus Damiao,}
\author[c]{H. Nogima,}
\author[c]{A. Santoro,}
\author[c]{S. Fonseca De Souza,}
% ...

\affiliation[a]{Ghent University, Dept. of Physics and Astronomy, Proeftuinstraat 86, B-9000 Ghent, Belgium}
\affiliation[b]{Centro Brasileiro Pesquisas Fisicas, R. Dr. Xavier Sigaud, 150 - Urca, Rio de Janeiro - RJ, 22290-180, Brazil}
\affiliation[c]{Dep. de Fisica Nuclear e Altas Energias, Instituto de Fisica, Universidade do Estado do Rio de Janeiro, Rua Sao Francisco Xavier, 524, BR - Rio de Janeiro 20559-900, RJ, Brazil}
% ...

\emailAdd{amrutha.samalan@cern.ch}

\abstract{During the upcoming High Luminosity phase of the Large Hadron Collider (HL-LHC), the integrated luminosity of the accelerator will increase to 3000 fb$^{-1}$. The expected experimental conditions in that period in terms of background rates, event pileup, and the probable aging of the current detectors present a challenge for all the existing experiments at the LHC, including the Compact Muon Solenoid (CMS) experiment. To ensure a highly performing muon system for this period, several upgrades of the Resistive Plate Chamber (RPC) system of the CMS are currently being implemented. These include the replacement of the readout system for the present system, and the installation of two new RPC stations with improved chamber and front-end electronics designs. The current overall status of this CMS RPC upgrade project is presented.}


\arxivnumber{} 

\collaboration[c]{on behalf of the CMS collaboration}

\proceeding{22nd International Workshop on Radiation Imaging Detectors\\
 27 June 2021 to 1 July 2021\\
}

\begin{document}
\maketitle
\flushbottom
\section{Upgrade of CMS sub-detector systems: RPCs}
\label{sec:intro}
The Large Hadron Collider (LHC) at European Council for Nuclear Research (CERN) was originally designed to operate with an instantaneous luminosity of 10$^{34}$cm$^{-2}$s$^{-1}$ and integrated luminosity on the order of 300 fb$^{-1}$. Already in the years 2015–2017, the maximum luminosity exceeded the design value by a factor of more than 1.7, which brought the CMS subsystems close to their rate capability limits \cite{adolphi2008cms}. During High Luminosity LHC (HL-LHC) operation, the instantaneous luminosity will be increased to 5$\times$10$^{34}$cm$^{-2}$s$^{-1}$, five times the machine’s original design value, and the ultimate performance of the HL-LHC would enable the collection of 400 to 450 fb$^{-1}$ of integrated luminosity per year. This increase in luminosity at the HL-LHC has many implications for the CMS sub-detector systems. It will increase the average pileup, which will result in worse background conditions, higher trigger rates, more difficult event reconstruction, and accelerated aging of components. To guarantee that the performance of the sub-detectors will be maintained during the HL-LHC running, CMS has started implementing different upgrades of the present systems, including extensions of systems with new chambers.

\begin{figure}[h]
\centering
\includegraphics[width=20pc]{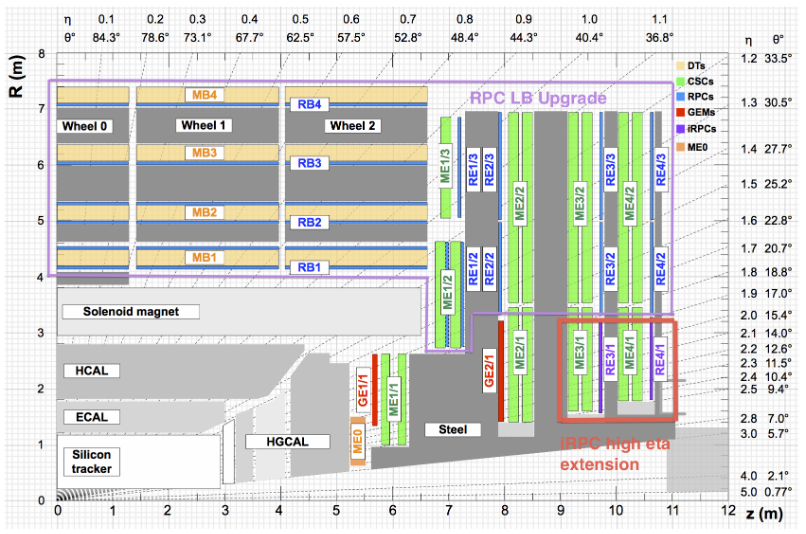}
\caption{A quadrant of the CMS experiment. The iRPCs will be installed in the region enclosed by the red box \cite{CERN-LHCC-2017-012}.}
\label{eta}
\end{figure}

The current RPCs at CMS are double-gap chambers operated in avalanche mode with a high electric field. With an excellent intrinsic time resolution of about 1.5 ns, they are mainly used for accurate timing and fast triggering. Two main upgrades are foreseen for the present RPC system.  While the RPC chambers are capable of operating until the end of Phase-2, the link system, which connects the front-end board to the trigger processors, must be exchanged, as is explained in section \ref{sec:linksys}. In addition, to extend the RPC coverage in the high pseudorapidity ($\eta$) region, an improved version of the already existing RPCs, referred to as iRPCs, will be installed in the forward region on the 3rd and 4th endcap disks. Figure ~\ref{eta} shows the region where these new RPCs will be placed (marked by the red box)~\cite{CERN-LHCC-2017-012}. New low noise front-end electronics have been introduced for the iRPC readout. Several prototype chambers have been built to validate the new electronics and to verify the production technology as well as the detector performance and longevity. The present status of the iRPC project is summarized in this paper.
\vspace{-1.2 cm}
\section{RPC Link System upgrade}
\label{sec:linksys}
In the CMS experiment, the RPC chambers are read out, controlled, and monitored through the Link
System, which consists of 1592 electronics boards divided in to two kinds: Link Boards (LBs) and Control Boards (CBs). LBs work either as a Master or Slave LB. The present Link System has been running well for more than 13 years. However, after the third LHC Long Shutdown (LS3), it is expected that aging effects will start to appear \cite{boghrati2021cms}. In addition, to deal with the higher data rate resulting from the HL-LHC, an increase in the data transmission band width is required, which is not possible using the present link system. Furthermore, most of the application specific integrated circuit (ASIC) chips of the link system will be outmoded by that time, and system maintenance will become a challenging issue. 
As such, the current Link System will be replaced by a new, upgraded system.

For the new system, the 28-nm Xilinx Kintex-7 was chosen as the best FPGA option. It offers a large amount of digital resources and supports high speed data transmission together with the ability to resist radiation \cite{boghrati2021cms}. Hence, an industrial version of the Kintex 7K160T was selected. To improve the time resolution of the muon chamber hits to the order of the RPC intrinsic time resolution, a 96-channel Time to Digital Converter (TDC) based on a combination of the logic elements and ISERDES primitives, was developed and implemented into the Kintex-7. Each TDC channel consists of 16 divisions or bins. The time resolution of each bin is one sixteenth of each bunch crossing, i.e. 1.56 ns. In a comparative study between the emulated and expected channel widths, the time resolution of each bin was compatible with the expectation based on the firmware emulation. Moreover, the differential and integral non-linearity errors of the TDC were well below 0.006 LSB and 0.01 LSB, respectively. Apart from this, the GTX transceivers of the Kintex-7 were used as a fundamental element to increase the speed of the data transmission line.

\section{Extension of the RPC system}
\begin{figure}[h]
\centering
\begin{minipage}{16pc}
\includegraphics[width=14pc]{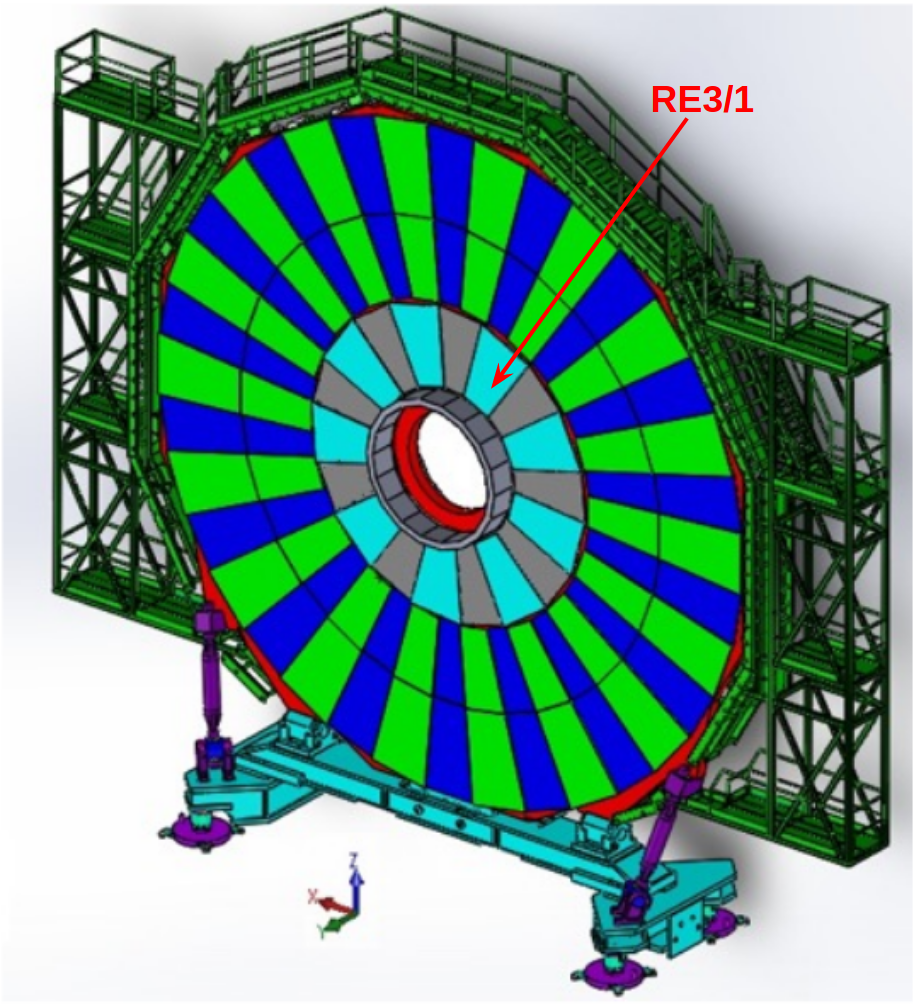}
\caption{Schematic view of the RE3/1 chambers mounted on the endcap disks \cite{CERN-LHCC-2017-012}.}
\label{mount}
\end{minipage}\hspace{1pc}
\begin{minipage}{18pc}
\includegraphics[width=16pc]{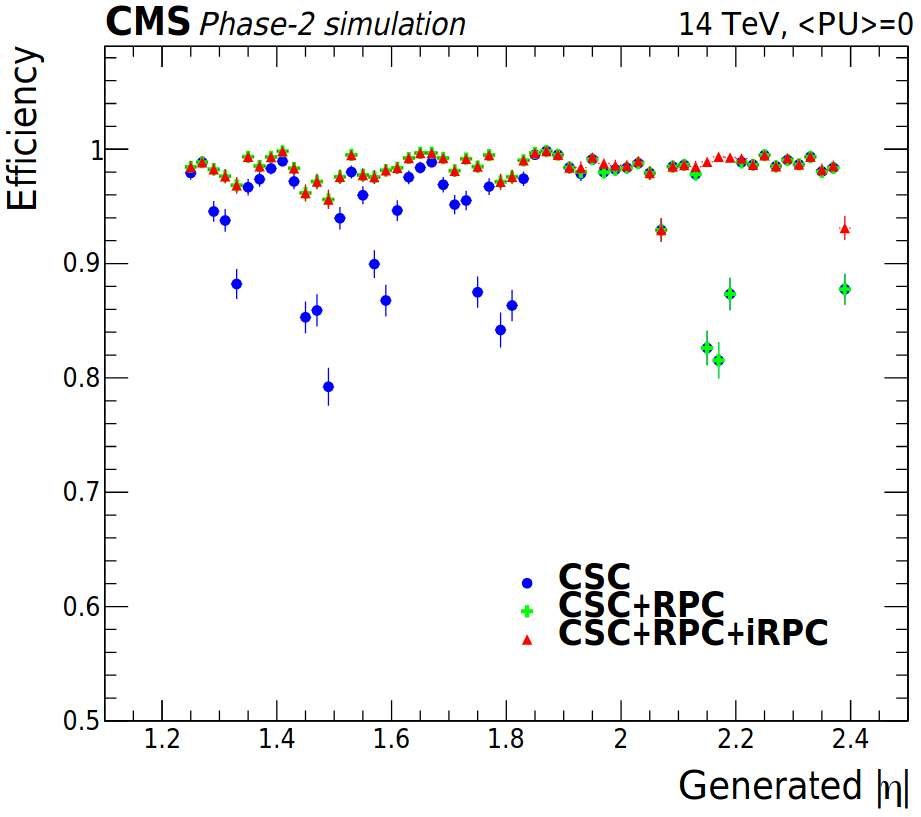}
\caption{Simulated comparison between the single muon trigger efficiencies with and without the RPC information, as a function of $|\eta|$ \cite{CERN-LHCC-2017-012}.}
\label{trig}
\end{minipage} 
\end{figure}

During the upcoming HL-LHC operation, the two endcap regions (high $\eta$ region) of the CMS experiment where the magnetic field is lowest will face high muon and background rates. For efficient muon triggering and offline particle identification as well as reconstruction, the measurement of a sufficient number of muon hits per track is required. The present muon system of the CMS experiment consists of 1056 RPC detectors that cover the region up to $|\eta|$ = 1.9. To increase the redundancy and robustness of the muon system, iRPCs (RE3/1 and RE4/1) will be installed in the forward region (on station 3 and station 4), extending the RPC coverage from $\eta$ = 1.9 to $\eta$ = 2.4. Figure \ref{mount} shows a schematic view of the RE3/1 chambers mounted on the endcap disks. These new chambers will complement the already existing Cathode Strip Chambers (CSC) ME3/1 and ME4/1. 

The major motivations behind upgrading the RPC system with iRPCs are: (i) by complementing the CSC stations, there will be an enhancement in the local muon measurement by adding track hits and by increasing the lever arm and (ii) the intrinsic time resolution will be improved (by a factor of $\sim$2) when the measurements of iRPCs are combined with those of the existing CSC chambers. This will improve the  background hit rejection, identification, and reconstruction of slowly moving heavy-stable-charged particles. In addition, unlike RPCs, iRPCs will measure the time difference between the signals at both ends of each readout strip, offering a better spatial resolution on the order of a few cm along the strip direction (non-bending projection). Profiting from this improved time and spatial resolutions of the additional iRPC hits and combining them with the CSC data, it will be possible to resolve the issue of low p$_T$tracks being misidentified as having high transverse momentum. Moreover, including iRPC hits in the trigger primitive stub-finding algorithm increases the triggering efficiency by identifying muon hits that the CSCs cannot detect at some values of $|\eta|$ due to the presence of high voltage (HV) spacers inside the chambers. The impact of adding the RPC and iRPC hit information to the single muon trigger is shown in Fig. \ref{trig}. A clear improvement in the trigger efficiency of around 15$\%$ can be observed between $|\eta|$ = 2.1 and 2.2.

\section{Technical specifications of iRPCs}
\begin{figure}[h]
\centering
\includegraphics[width=13pc]{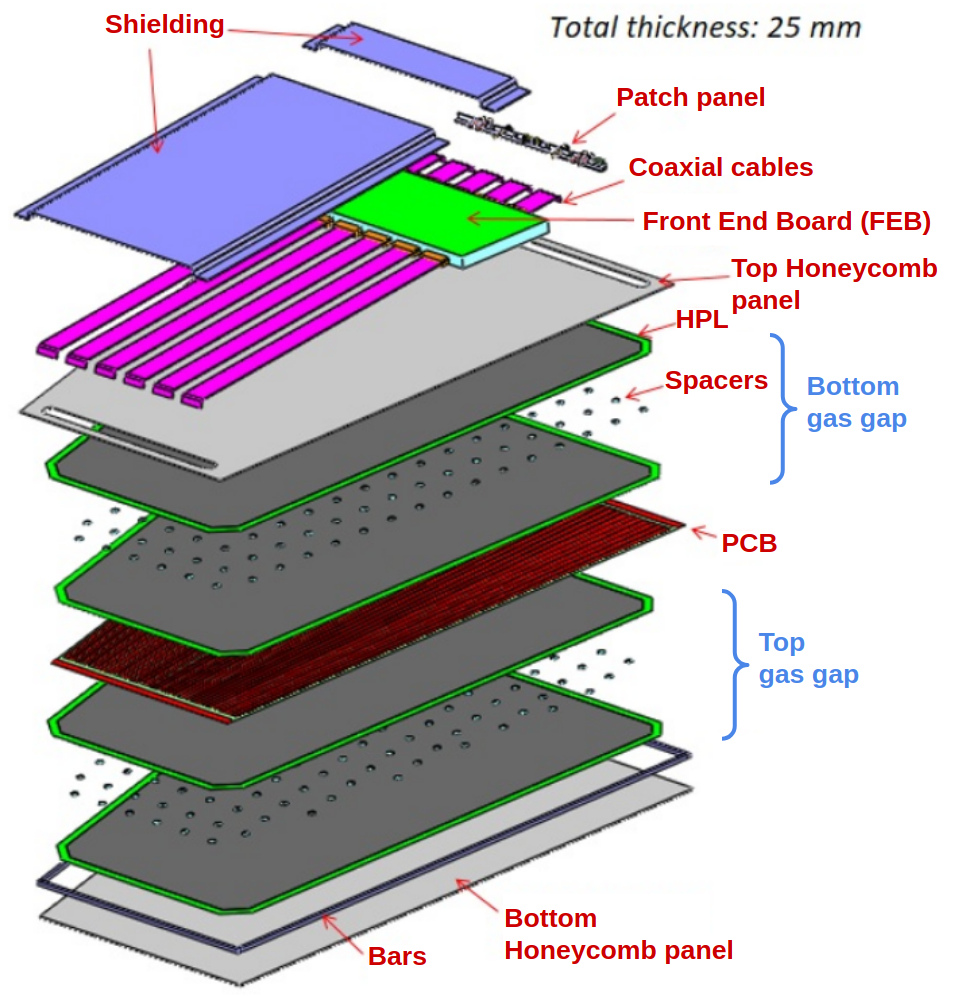}
\caption{Schematic view of an iRPC detector.}
\label{irpc}
\end{figure}

To operate under HL-LHC conditions while maintaining a hit efficiency above 95$\%$, an improved version of the technical RPC design is introduced for the new iRPCs. From the layout point of view, the chambers are similar to the existing wedge-shaped RPC detectors with radially oriented readout strips placed between the gaps. An iRPC chamber consists of two gaps, i.e. top and bottom, each made of two High Pressure Laminate (HPL) electrodes coated with a thin graphite resistive layer. Unlike the present RPC design, the strips are integrated in a large trapezoidal printed circuit board (PCB) made of two parts. Each PCB comprises 96 readout strips in total, with a pitch of about 12.3 mm in the high radius (HR) section and about 6 mm in the low radius (LR) section. 
Two Front End Boards (FEBs) containing the front-end electronics are directly plugged into the PCB. Compared to the existing RPC system, the new readout scheme offers a better spatial resolution and reduces the number of electronic channels by 60$\%$ \cite{meola2021towards}. The new layout scheme also has the additional advantage of facilitating the cabling layout and the chamber construction because the strip signals have to be extracted from only two sides of the chamber and not from different partitions as in the present RPCs. To reduce possible aging effects and to improve the rate capability \cite{aielli2016improving}, the thickness of the electrodes as well as the gas gaps are reduced from 2 mm to 1.4 mm. A reduction in the efficiency is avoided by amplifying the signal and increasing the sensitivity of the front-end electronics system five times compared to the existing system \cite{lee2018study}. By improving the front-end electronics and simultaneously reducing the gas gap thickness, the avalanche charge is effectively reduced, and hence the rate capability and chamber longevity are enhanced. Following this new layout, the high-voltage working point of the iRPC detector is also reduced from 9.5 to 7.1 kV. An exploded view of an iRPC chamber is shown in Fig. \ref{irpc}.
\section{New front-end electronics for iRPCs}
With the new layout of the iRPCs, the amount of charge deposited by a charged particle passing through the detector is reduced compared to the present CMS RPCs. The front-end board (FEB) of the present system does however not support detecting the lower charges without affecting the detector performance. Therefore, a new FEB equipped with low noise front-end electronics that can detect signals with a charge as low as 10 fC was designed and developed. It is fast and reliable and has the ability to survive in the high radiation environment. The front-end electronics of the iRPCs is based on the Petiroc ASIC, which is a 32-channel ASIC using a broad band fast preamplifier and a fast discriminator in SiGe technology \cite{fleury2013petiroc}. Its overall bandwidth is 1 GHz with a gain of 25, and each channel provides a charge measurement and a trigger output that can be used to measure the signal arrival time. The new PCBs offer the possibility to readout each strip from both ends, yielding two signals per pitch.
Profiting from the excellent time resolution (20–30 ps \cite{Lagarde_2016}) of the electronics and by measuring the arrival time difference of the two strip ends of the 2D readout scheme, it is able to determine the position along the strip with good resolution of about 200 ps or 1$\cdot$7 cm \cite{electronics}.

\begin{figure}[h]
\centering
\centerline{\includegraphics[width=24pc]{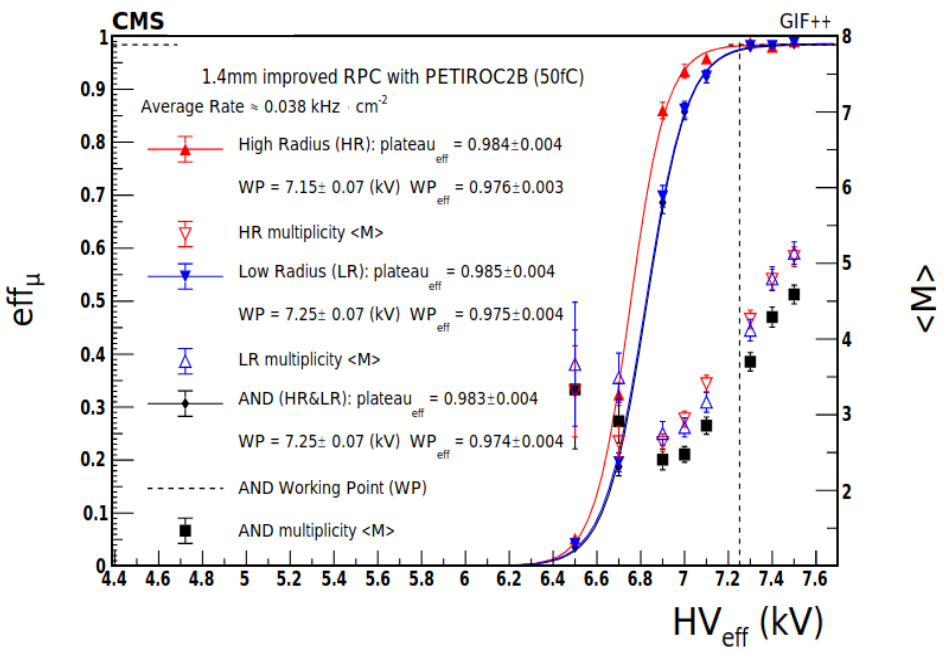}}
\centerline{\includegraphics[width=24pc]{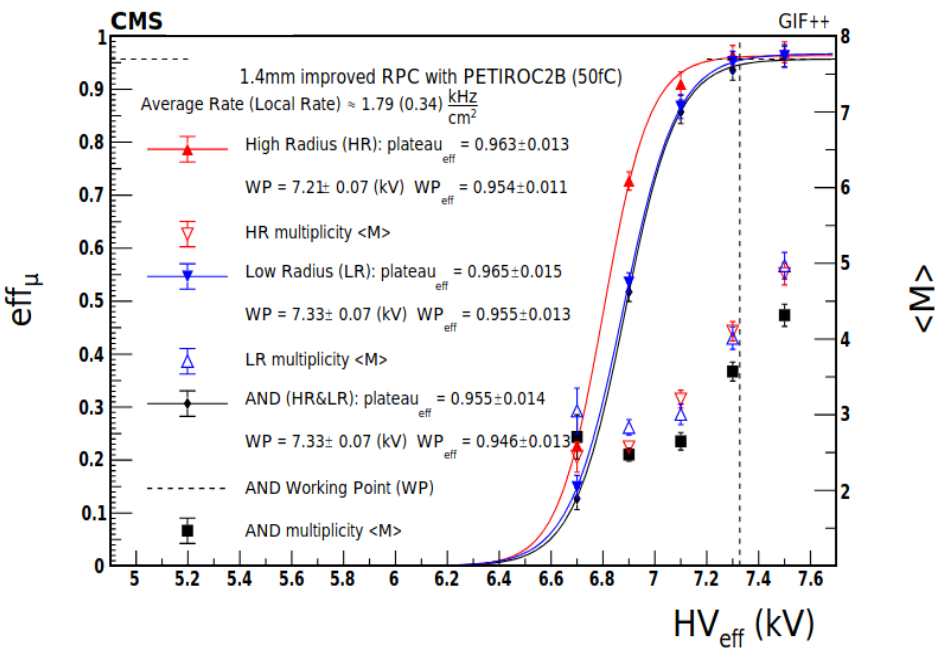}}
\caption{Efficiency versus effective HV at a low irradiation (top) and high irradiation (bottom). The hit multiplicity per event is also reported \cite{shchablo2021front}.}
\label{feb}
\end{figure}

To validate the performance of iRPCs with the new front-end electronics, different studies with the proposed FEB were conducted using a muon beam under different irradiation conditions at the CERN Gamma Irradiation Facility (GIF$++$) \cite{guidanew}. Although the GIF++ source produced a uniform irradiation all over the chamber during the test, the presence of a lead shield resulted in different irradiation fluxes over different regions of the iRPC. This variation allowed the study of the impact of the electronics dead time on the iRPC efficiency under a high flux counting rate up to 2 kHz$\cdot$cm. The top (bottom) plot in the Fig. \ref{feb} shows the efficiency measured at a low (high) irradiation with an average rate of $\sim$ 0.038 (1.790) kHz$\cdot$cm$^{-2}$ with respect to the effective high voltage (HV$_{eff}$) on the gaps and also the variation of the hit multiplicity (<M>) according to the HV$_{eff}$ \cite{Paolucci_2013}. Even at the high background rate, around a factor 3 higher than that expected in the HL-LHC period, efficiencies above 95$\%$ were obtained at the chamber working point. The light red (dark black) line shows the efficiency with the signal detected on the HR (LR) side.  The efficiencies were measured at a position close to the HR side of the chamber. The slight difference in efficiencies observed between the HR and LR sides is due to the signal loss through propagation along the strip. The PCBs equipped in the chamber used for this test were made of Flame Retardant (FR4) material with a dielectric constant of about 4.3. The loss of efficiency arising from the signal attenuation when propagating along the strips is expected to be reduced by developing the next version of the strip panels with EM888 material of dielectric value 3.7 \cite{shchablo2021front}.

\section{iRPC demonstrators}
To study the detector behaviour under real LHC conditions such as high background rate, noise, magnetic field, etc.; to validate the iRPC FEB and backend electronics; and to integrate the new RPC stations into the CMS detector control software and data acquisition system, a number of iRPC demonstrator chambers will be installed in CMS during the ongoing, 2nd LHC Long Shutdown. This project will also lead to acquiring installation expertise in the CMS-RE3/1 and RE4/1 regions. The production center of HPL panels for iRPCs as well as RPCs is the Riva Laminati company (Milan, Italy), and the gaps are produced at the Korean Detector Laboratory. Eight demonstrator chambers (5 RE3/1 and 3 RE4/1) were assembled in July 2021 at Ghent University (Belgium).  Gap-level as well as chamber-level Quality Control tests (QCs) were performed on the site following standardized protocols.
The gap level QC includes visual inspection of the gaps, verification of the spacer conditions and gas tightness, and dark current measurements. The gas tightness test and spacer test for iRPCs are conducted at an overpressure of 15 hPa above atmospheric pressure.
\begin{figure}\centering
\subfloat[Leak test]{\label{a}\includegraphics[width=.41\linewidth]{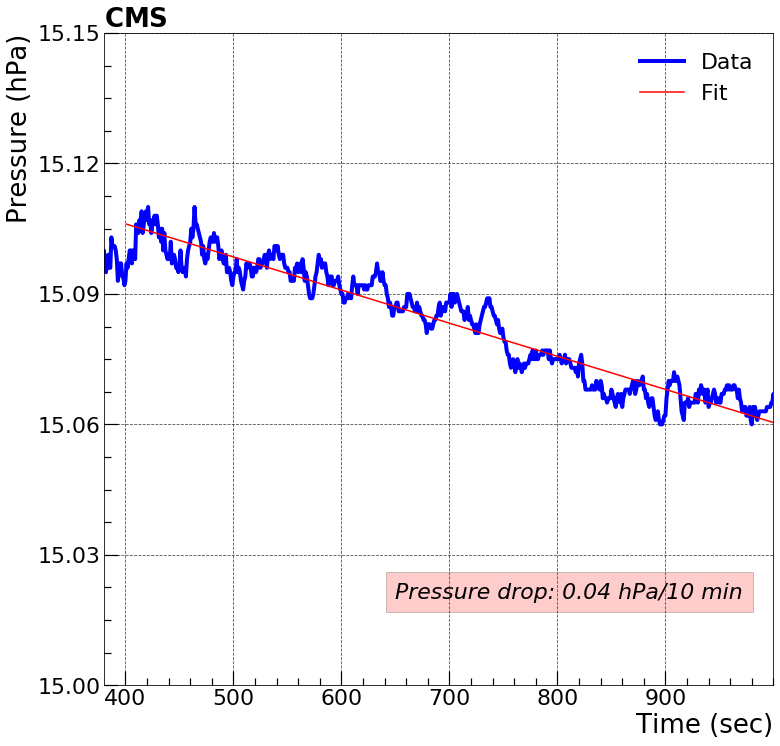}}\hfill
\subfloat[Spacer test]{\label{b}\includegraphics[width=.41\linewidth]{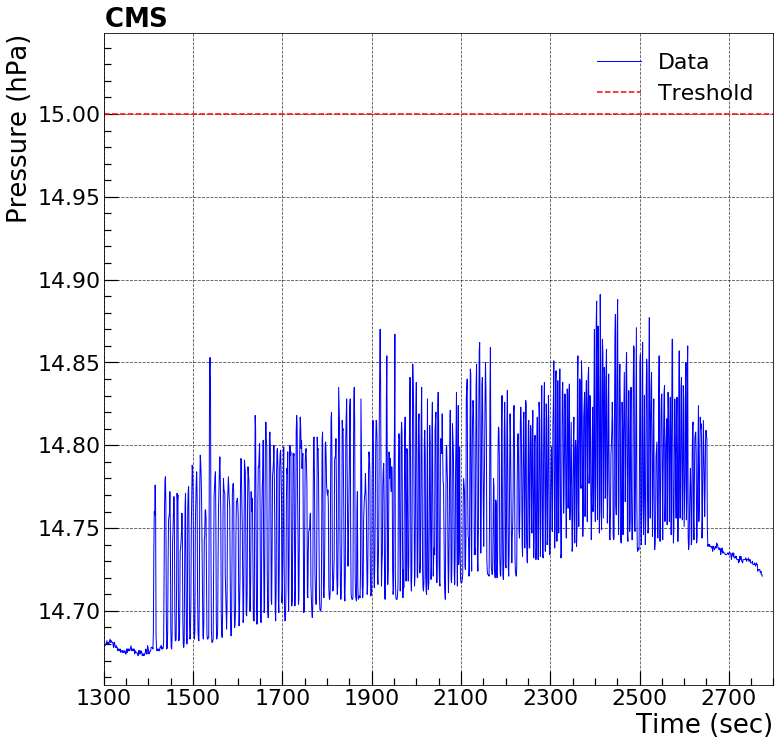}}\par 
\subfloat[Dark current test]{\label{c}\includegraphics[width=.41\linewidth]{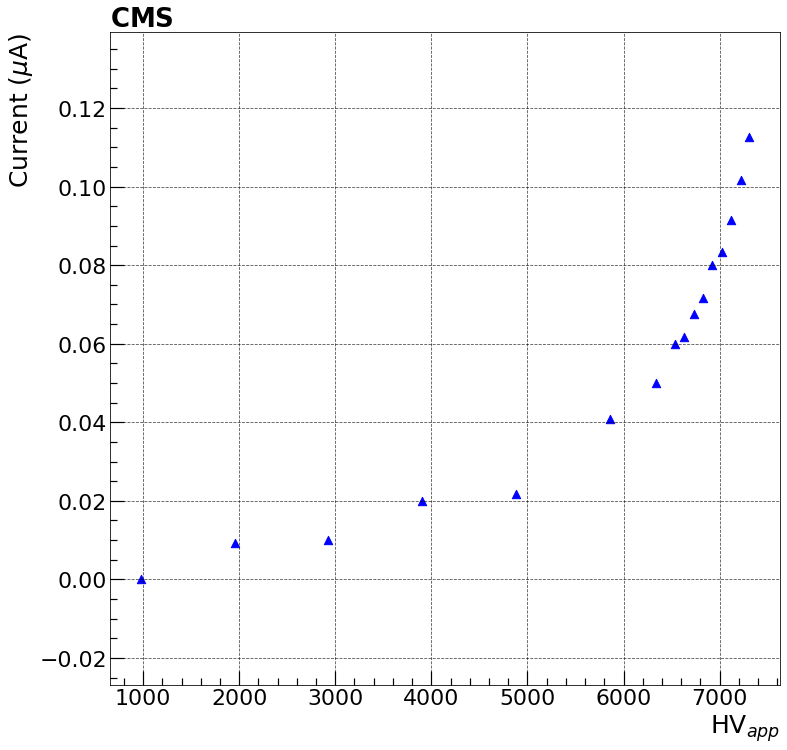}}
\caption{The gap level QC plots of the CMS iRPC demonstrator chamber.}
\label{qc}
\end{figure}
Figure \ref{qc} shows the plots of the gap-level QC tests performed for a demonstrator gap. In the gap-level QC test, a positive pressure of about 15 hPa is applied to each gap and the pressure drop is measured as a function of time, as shown in the plot \ref{a}. The leak rate, i.e. slope of the linear fit (light red line) of the data (dark blue), for the acceptable gap should be less than 0.04 hPa/min. The plot \ref{b} shows the monitored pressure for the spacer test. The dark blue trend shows the data and the horizontal light red line in the upper part of the plot shows the threshold for considering the spacer as a popped up one. Each spike in the plot corresponds to a spacer in the gap. The transitions show that the spacers are in good condition. For the HV test, the dark current is monitored at 16 HV points. The plot \ref{c} shows the variation of the dark current with respect to the applied HV, which is in agreement with the expected trend. As shown in the plot, at 5 kV the iRPC gap becomes fully ohmic and the working point is at 7.1 kV.

After the assembly, a visual inspection of the chamber mechanics is performed and the gas leak tests and dark current tests are repeated to verify the status of the gaps. All chambers are shipped to the RPC laboratory at CERN and the remaining QC, which include the long term chamber stability check and the cooling circuit, electronics, and cosmic muon tests are ongoing. Four qualified chambers out of the eight are scheduled for installation on the CMS endcaps in November 2021.
\section{Summary}
As part of the effort to  maintain the robustness and redundancy as well as the identification and reconstruction capabilities of the CMS muon system during the HL-LHC period, several upgrades are being implemented to the CMS RPC system. The present RPC Link System is being upgraded to make the RPC system robust against background noise and to fully exploit the intrinsic  time  resolution on the order of 1.5 ns of  the  iRPC chambers. To extend the current RPC pseudorapidity coverage up to $|\eta|$ = 2.4, 72 new iRPCs will be added to the 3rd and 4th endcap stations. An updated technical layout is proposed for the iRPCs in which the thickness of the gas gaps and the RPC electrodes are reduced from 2 to 1.4 mm, to offer a short recovery time for the avalanche charge through the RPC electrodes, enhancing the chamber rate capability and reducing possible effects of detector aging. The RPC upgrade also includes the replacement of the present readout system. The performance of the new front-end electronics was validated at the Gamma Irradiation Facility at CERN, i.e. an efficiency above 95$\%$ was measured at a high particle rate of 1.79 kHz$\cdot$cm$^{-2}$, which proves the capability required for reliable operation under HL-LHC conditions. As part of the iRPC demonstrator project, eight chambers built at Ghent University are undergoing detailed performance studies at CERN. The installation of these iRPC demonstrator chambers in CMS is scheduled for November 2021.

\acknowledgments
The corresponding author would like to acknowledge the support of the FWO-Flanders (Belgium, project no. I002319N), and to express her sincere thanks to the iWoRiD-2021 conference committee for organizing the workshop.


\bibliographystyle{unsrt}

\bibliography{iRPC_iWoRiD}
\end{document}